\documentclass[letterpaper,twocolumn,10pt]{article}
\usepackage{usenix-2020-09}

\usepackage[most]{tcolorbox}
\usepackage{booktabs}

\usepackage{graphicx}
\usepackage{multirow}
\usepackage{amssymb} %
\usepackage{amsfonts} %
\usepackage{calligra} %
\usepackage{mathtools}
\usepackage{amsthm} %

\usepackage{pgfplots}
\usepackage{pgfplotstable}
\usepgfplotslibrary{statistics}

\makeatletter
\let\MYcaption\@makecaption
\makeatother
\usepackage[font=footnotesize]{subcaption}
\makeatletter
\let\@makecaption\MYcaption
\makeatother

\usepackage{comment}
\usepackage{listings}

\usepackage{circledsteps}

\usetikzlibrary{patterns}

\usepackage{blindtext}

\usepackage{acronym}

\usepackage{xcolor, colortbl}

\usepackage{float}
\usepackage{textcomp}

\usepackage{algorithm}
\usepackage{algorithmic}

\usepackage{url}

\usepackage[all=normal,paragraphs=tight, floats=tight,indent=tight]{savetrees}
\usepackage{pifont}
\usepackage{framed}
\usepackage{soul}
\usepackage{multirow}
\usepackage{array}
\newcolumntype{L}[1]{>{\raggedright\let\newline\\\arraybackslash\hspace{0pt}}m{#1}}
\newcolumntype{C}[1]{>{\centering\let\newline\\\arraybackslash\hspace{0pt}}m{#1}}
\newcolumntype{R}[1]{>{\raggedleft\let\newline\\\arraybackslash\hspace{0pt}}m{#1}}
\newcolumntype{H}{>{\collectcell\lstinline}l<{\endcollectcell}}
\usepackage{url}

\newcommand{\cmark}{\ding{51}}%
\newcommand{\xmark}{\ding{55}}%

\usepackage{enumitem}

\newcounter{question}
\newenvironment{q_and_a}{\vspace{0.5em}
\begin{itemize}[noitemsep,topsep=0pt,parsep=0pt,partopsep=0pt,font=\bfseries]\hrule
}{\end{itemize}\vspace{0.5em}\hrule\vspace{0.5em}}

\newcommand{\question}[1]{\vspace{0.5em}\stepcounter{question} \item[Q\thequestion.] \textit{#1}}
\newcommand{\answerT}[3]{\item[#2 A.] ($t=#1$) ``#3''}
\newcommand{\answerM}[3]{\item[#2 A.] (#1) ``#3''}
\newcommand{\answer}[2]{\item[#1 A.] ``#2''}
\include{sec/preamble/preamblelistingsconf}
\acrodef{LLM}[LLM]{large language model}
\acrodef{ML}[ML]{machine learning}
\acrodef{PID}[PID]{Proportional-Integral-Derivative}
\acrodef{ICS}[ICS]{Industrial Control System}

\newcommand{\davinci}{\texttt{code-davinci-001}}
\newcommand{\cushman}{\texttt{code-cushman-001}}
\newcommand{\babbage}{\texttt{code-search-babbage-code-001}}

\begin{document}

\date{}

\title{\Large \bf Pop Quiz! Can a Large Language Model Help With Reverse Engineering?}

\author{
{\rm Hammond Pearce}\\
New York University\\
hammond.pearce@nyu.edu
\and
{\rm Benjamin Tan}\\
University of Calgary\\
benjamin.tan1@ucalgary.ca
\and
{\rm Prashanth Krishnamurthy}\\
New York University\\
prashanth.krishnamurthy@nyu.edu
\and
{\rm Farshad Khorrami}\\
New York University\\
khorrami@nyu.edu
\and
{\rm Ramesh Karri}\\
New York University\\
rkarri@nyu.edu
\and
{\rm Brendan Dolan-Gavitt}\\
New York University\\
brendandg@nyu.edu
} %

\maketitle

\begin{abstract}

Large language models (such as OpenAI's Codex) have demonstrated impressive zero-shot multi-task capabilities in the software domain, including code explanation. 
In this work, we examine if this ability can be used to help with reverse engineering. Specifically, we investigate prompting Codex to identify the purpose, capabilities, and important variable names or values from code, even when the code is produced through decompilation.  
Alongside an examination of the model's responses in answering open-ended questions, we devise a true/false quiz framework to characterize the performance of the language model. We present an extensive quantitative analysis of the measured performance of the language model on a set of program purpose identification and information extraction tasks: of the 136,260 questions we posed, it answered 72,754 correctly. A key takeaway is that while promising, LLMs are not yet ready for zero-shot reverse engineering.  %

\end{abstract}

\section{Introduction}

\acp{LLM} such as GitHub Copilot~\cite{github_github_nodate}, OpenAI's Codex~\cite{chen_evaluating_2021,openai_openai_2021}, and AI21's Jurassic-1~\cite{lieber_jurassic-1_2021} are increasingly being promoted for use within the software development domain. 
Such models are built using \ac{ML} over vast quantities of unstructured text, including websites, books, and open source codes. This enables them to produce `completions' given some input prompt made up of code and comments (documentation). 
While these models are primarily being advertised for code creation, other use-cases have been highlighted, including programming language translation, bug repair, and code summarization and explanation~\cite{ai21_discover_nodate,openai_examples_nodate,pearce_can_2021}, all in a \textbf{zero-shot} setting.

Reverse engineering is the process by which we try to understand how an existing artifact works, for both malicious or defensive purposes. In this human-intensive process, reverse engineering relies on knowledge and experience. 
As such, \textbf{given that recent \acp{LLM} have been trained on vast quantities of code across a dizzying array of applications~\cite{chen_evaluating_2021}, to what extent can they help us reverse engineer software? }
To answer this question, we investigate the usefulness of an ``out-of-the-box'' \ac{LLM} in explaining code, including that which comes stripped and decompiled. 
In contrast to prior work where models are trained for a specific reverse engineering task (as we discuss in \autoref{sec:relatedwork}), general \acp{LLM} offer us the chance to ``ask'' open-ended questions about code. %

The purpose of reverse engineering software in a cybersecurity context is to identify valuable information, such as the functionality and purpose of functions/data to exploit or repair. 
The ability of an \ac{LLM} to provide explanations about code arises from the model learning to recognize functional intent in open source software captured by code artifacts such as function and variable names or code comments~\cite{misra_is_2020}. A model, which is given a code sample that is structurally familiar to its training data, for example, can perhaps make an educated guess about what it does. 

To examine whether a leading \ac{LLM} has `learned' to recognize program purpose and other information, even in the absence of comments (e.g., decompiled source code), we consider the following questions: to what extent can it predict the purpose of a given code or extract requested information (e.g., variables' roles and values)? Can it still answer questions about code after we strip out meaningful function/variable names? 
From seeking answers to these questions, \textbf{our main contributions are}:
(i) a quantitative analysis framework of the performance of LLMs for the purpose identification and information extraction tasks and (ii)
an experimental study of the LLM's performance for the purpose of identification and information extraction tasks on a range of test programs drawn from real-world scenarios including malware and industrial control systems.

\section{Prerequisites: Motivation and Background}

\subsection{Reverse Engineering}
There are numerous investigative aspects in the domain of cybersecurity where both defenders and adversaries aim to collect information about software. 
This includes manual analysis of software to understand both functionality and form---what the software does and how it tries to do it---so that they gain information about assets and weaknesses in a design. 
While decompilation tools such as Ghidra~\cite{national_security_agency_ghidra_nodate} exist to transform binaries into source code, reverse engineering requires considerable human-insight-in-the-loop and effort to make sense of that code. 
Intuitively, one needs to be familiar with different application domains and software design approaches to recognize artifacts of interest in an unseen code. 

A seminal paper by Chikofsky and Cross defines \textit{Reverse Engineering} as the process of identifying the components of a system and their interrelationships, and generating representations of the system in another form or at a higher level of abstraction~\cite{chikofsky_reverse_1990}. They clarify that the reverse engineering process is an examination, not a process of construction, change, or replication.  Thus reverse engineering encompasses various life cycle stages, starting from the current implementation---recapturing or recreating the design and deciphering the requirements implemented by the subject system. 

In its broadest sense, reverse engineering refers to reproducing all the information needed in order to comprehend a given program: i.e., the program's purpose, its function, why it does it, how it accomplishes its function, etc. As this represents a much wider range of information than typically found in software representations or codes, we will limit our analysis to recovering the kinds of details that are found within software itself: in comments, variables, hard-coded/preset values of variables, naming schemes, etc.

\subsection{Example Domains}
\label{sec:motivation-malware-re}
Let us consider two application domains where we are interested in gleaning author intent from binaries---reverse engineering of malware and industrial control systems. 

Malware is designed to exploit one or more vulnerabilities in a system to perform unwanted behaviors, including denial of service or violating privacy/secrecy. 
Reverse engineering malware provides insights into the specific mechanisms that it employs for attack and evasion, leading to insights that can be used for implementing detection or mitigation mechanisms. 
Reverse engineering can also provide clues as to the provenance of malware, such as when we identify similarities (or fingerprints) in the way malware is designed~\cite{alrabaee_survey_2022}. 

\label{sec:motivation-industrial-re}
Reverse engineering of code is also relevant in the \ac{ICS} domain.
In legacy ICS devices, extracting information from source/binaries such as the code purpose, mathematical equations, code structures, variable names, and parameter values can be vital in maintenance of legacy systems and in re-implementation on newer devices. With configuration/logic changes introduced over several years by multiple operators, such information might not be properly documented or readily available. %
While compiling mathematical equations into binary code to execute on diverse controller hardware is straightforward, \textit{reversing} this process is very difficult. However, this process may be vital to be able to reconstruct control logic in a mathematical form that can be analyzed, modified, or re-implemented. Reverse engineering \ac{ICS} can also be pivotal when analyzing \textit{cyberattacks} on those systems, especially given that increasing complexity and connectivity of modern-day \acp{ICS} increases their exposure to such attacks. For instance, it may be necessary to perform reverse engineering of the deployed software to determine nature of changes (e.g., modifications of control logic in \ac{ICS}) introduced by malicious third parties. %

Since real-world codes in both example domains are typically written in C (or are compiled to C from high-level representations such as ladder logic in ICS devices), we focus on C source code in this paper. Focusing on C codes also enables us to explore the LLM performance variations between source and decompiled code since decompiled C code tends to be more dissimilar to the source C code (unlike Java and Python where decompiled code is much more similar to the original source), therefore making decompiled C a more challenging scenario for the LLM.

\subsection{Can LLMs Help?}
In both these domains, as well as others, reverse engineering requires considerable domain expertise and is labor-intensive. 
Hence, we are motivated to explore approaches that might be able to help us understand elements of code, especially where the code might be unfamiliar to a human (such as after decompilation/disassembly).  
Given the emergence of \acp{LLM} such as OpenAI's Codex, the models' exposure to a wide variety of code (through training on open source software), and their apparent ability to multi-task ``out-of-the-box'', such as  code summarization~\cite{openai_examples_nodate}, we were curious to see if we can ``quiz'' an \ac{LLM} to get useful information about code. 

\begin{figure}[t]
    \centering
    \includegraphics[width=\linewidth]{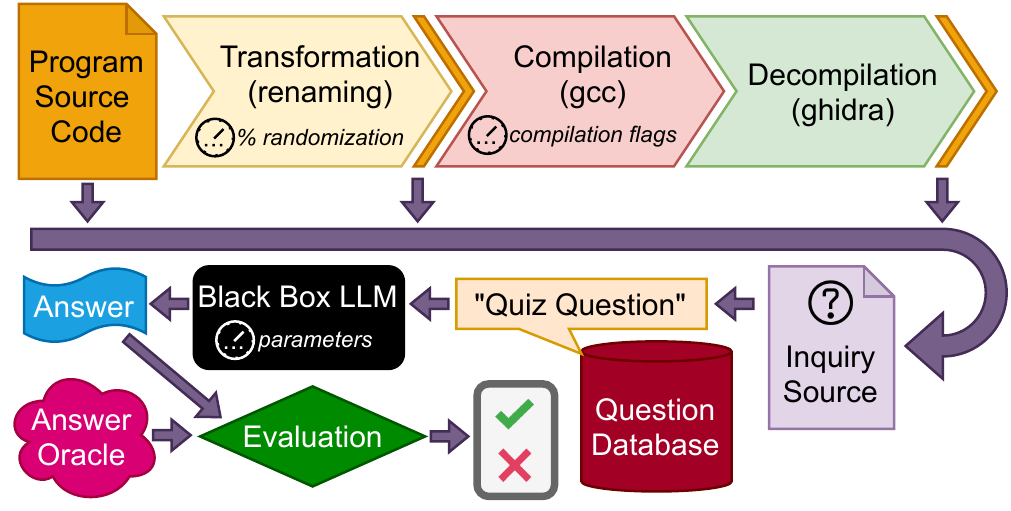}
    \caption{Reverse engineering inquiry evaluation system.}
    \label{fig:re-proposal}
\end{figure}

The general way in which one interacts with an \ac{LLM} is to provide an input prompt, such as selected lines of code from a program of interest.  
As part of the prompt, we can add a question about the functionality/role of artifacts in the code. 
In response, the model produces a sequence of tokens containing an answer to the provided question. 
Thus, our study explores the process of ``quizzing'' the model to get information about code, which we put together in an experimental framework as shown in \autoref{fig:re-proposal}.

\label{sec:llm-reversal-process}

\section{Pop Quiz! Initial Exploration}

Outside of reverse engineering, \ac{ML} has shown some promise in tasks such as code summarization (as we will discuss in \autoref{sec:relatedwork}). 
However, to the best of our knowledge, there are not yet standardized, universal benchmarks for quantifying a model's ability to assist in a ``reverse engineering'' context. 
Hence, we begin our foray by performing a series of informal experiments, drawing qualitative, empirical observations from our interaction with the \ac{LLM}. 
For this research, we use OpenAI's \davinci~\cite{chen_evaluating_2021} (also known as \textit{davinci-codex}) as the basis of our evaluation. This model was chosen primarily for the large token size it supports (4098 tokens compared with the more common limit of 2048 tokens in OpenAI \cushman\ and Jurassic J-1 models from AI21~\cite{ai21_discover_nodate}).
While we focus on OpenAI's Codex for experimental studies in this paper, several LLMs are available that provide summarization capabilities (e.g., GitHub's Copilot). The fact that such automated summarization and text/code prediction features are being built into multiple LLM platforms is indeed a major motivating factor making the study in this paper timely and widely relevant. 

We initially examine the ability for program information to be derived and extracted from a program's original source code, before applying transformations such as renaming variables and functions to random strings, and compilation and decompilation using \texttt{gcc} and Ghidra~\cite{national_security_agency_ghidra_nodate} to progressively emulate the challenges faced by reverse engineers.

\subsection{Small-Scale Experiments}
\label{sec:initial-experiments}
To begin, we design a small-scale experiment to explore two short programs within the OpenAI `playground' environment. 
These programs serve as a running-example throughout this paper. 
We hand-crafted a series of fairly open-ended questions about the code's functionality and variables to ask the model. 
We design the protocol using the approach laid out in the OpenAI examples for `Code Explain'~\cite{openai_examples_nodate}.
    We begin the prompt with a `language hint', which in this paper is always ``\texttt{//c}'', indicating that the program is in the C programming language. 
    The program is then added in its entirety. 
    We then add a section break (three double quote marks), before beginning the `Q\&A' section.

\subsubsection{A basic `malware' component}
\label{sec:initial-explore-malware}
\begin{figure}
\centering
\input{fig/listings/delete-listen}
\vspace{-3mm}
\caption{C `malware' \textbf{delete\_listen} which deletes files from directory paths received via open socket, presented in a `Q\&A' template for use with the \davinci{} \ac{LLM}.}
\vspace{-5mm}
\label{fig:lst:prog-delete-listen}
\end{figure}

The first program is listed in \autoref{fig:lst:prog-delete-listen} according to our template (pink line 1: language prompt, yellow lines 57-59: section break and `Q\&A' area) and relates to the malware-based motivation presented in \autoref{sec:motivation-malware-re}. This program, which could be part of a `ransomware'-type malware, opens a socket on the host computer, and then waits for folder paths to be received via that socket. Once received, it deletes all files from those folders.

We present the results of our manually-driven experimentation with this first program. Initially, we set all parameters for  \davinci\  
to its defaults (\texttt{temperature} $t=0.0$, \texttt{top\_p} $=1.0$). We add a stop sequence `Q.' to discourage the \ac{LLM} from generating text beyond the desired answer (otherwise, we observed that the model is chatty with a tendency to continue the `Q\&A' format by generating its own questions). We repeated the prompt with the \texttt{temperature} $t=1.0$. As authors we interpreted the answers and judged them as correct ``\cmark'', incorrect ``\xmark'', or somewhere in  between ``\textbf{?}''. Importantly, we \textbf{do not} present multiple questions, or build a sequence of question-after-question (leaving the previous answer intact). Each question is presented \textbf{in isolation} according to the template in \autoref{fig:lst:prog-delete-listen}.
\begin{q_and_a}
\question{What is the purpose of the above code?}
\answerT{0}{\cmark}{It is a server that can delete files in a directory.}
\answerT{1}{\xmark}{In a victim's perspective, the server will keep listening on the port, receiving connection requests, accepts theClient's request, accepts data, and either individually processes the data, with repeated i/o requests, or then checks its control flag and sends some amount of data to theclient, proceeding in a similar manner.}

\question{Where does the directory to delete files come from?}
\answerT{0}{\cmark}{It comes from the client.}
\answerT{1}{\xmark}{\foreignlanguage{russian}{рюшектится через извещение в слушаешей нуждающимся}\footnote{Translated by Google, this reads `ruffles through the notice in the listener to the needy'; when translated by a colleague fluent in Russian it is `meaningless---random words in a string.'}}

\question{How is the directory to delete files communicated to the server?}
\answerT{0}{\xmark}{The directory to delete files is communicated to the server by the client.}
\answerT{1}{\cmark}{The server waits for incoming requests through the server socket contained in a while loop that recieves messages and acts upon the message if it it is a delete message.}

\question{What variable stores the folder to delete files from?}
\answerT{0}{\cmark}{dirname}
\answerT{1}{\cmark}{dirname}
\question{What is the format of the message the client sends to the server?}
\answerT{0}{\cmark}{The client sends a string of the form "d<path>" to the server.}
\answerT{1}{?}{Header: Command + SENDNAME / Payload(optional): Contents of the file / Header: Command + LNAME / Payload(when used): directory name / : command + argument}

\question{Can the client shut down the server? If so, how?}
\answerT{0}{\cmark}{Yes, the client can shut down the server by sending a `q' message to the server.}
\answerT{1}{?}{Of course the client can shutdown the server, we just have to send q\textbackslash nExit code on client: n}

\question{What IP address and port does the server listen on?}
\answerT{0}{\cmark}{The server listens on the IP address 192.168.0.1 and port 8080.}
\answerT{1}{\xmark}{ip="127.0.0.1" port=8888}
\end{q_and_a}

We make several observations from this ``pop-quiz'' Q\&A session. Firstly, we can see that many of the \ac{LLM}'s answers are correct even though our program is somewhat complex (and contains no comments!). 
This appears to support %
the claim by OpenAI and the other \ac{LLM} designers that such models \textit{can} read and summarize code snippets---and also starts to add credence to our hypothesis that these \acp{LLM} might be able to aid in reverse engineering even in this zero-shot setting.

That said, although a number of the answers are correct, many are not. Several of the answers at the high temperature value are gibberish, e.g., the answer to Q1 and the Russian output in Q2. 
The high temperature setting was able to elicit the correct answers for Q3, whereas the low temperature's answer is insufficient (it largely just restates the question). 
This indicates that some tuning will be required to identify the best model parameter settings for the \ac{LLM} in this use case. 

\begin{figure}[t]
\centering
\input{fig/listings/delete-listen-random}
\vspace{-3mm}
\caption{Randomized version of  \textbf{delete\_listen} from \autoref{fig:lst:prog-delete-listen}.}
\label{fig:lst:prog-delete-listen-random}
\vspace{-5mm}
\end{figure}

Code presented for reverse engineering might not always be as descriptive as the program listed in \autoref{fig:lst:prog-delete-listen}. 
Consider the case in \autoref{fig:lst:prog-delete-listen-random}, where every variable name, internal function name, \#define, address, and port are all randomized. 
Intuitively, it might seem that this should degrade \davinci's ability to interpret the code. 
However, when we ask some of the same questions, we observe that the model was still capable of producing correct answers (although perhaps at a lower incidence rate). Due to the lower success rate of the high-temperature answers, we now ask only with \texttt{temperature} $t=0$.
\begin{q_and_a}
\question{What is the purpose of the above code?}
\answer{\cmark}{To remove files in a directory.}

\question{Where does the directory to delete files come from?}
\answer{\cmark}{It comes from the client.}

\question{What variable stores the folder to delete files from?}
\answer{\cmark}{neklkzq}

\question{What is the format of the message the client sends to the server?}
\answer{\cmark}{The client sends a string of the form "d<path>" or "q" to the server.}
   
\question{What IP address and port does the server listen on?}
\answer{\cmark}{The server listens on port 7716 on the IP address 91.53.33.119.}

\end{q_and_a}

Impressively, \davinci{} appears to maintain the ability to extract information even after randomizing the code. 

For the next challenge, we consider the case where the original source is not available. 
Instead, we need to reverse engineer starting form a binary file compiled using \texttt{gcc -o1}. 
Using a tool like Ghidra, we recover functions such as \texttt{del\_in\_dir} and \texttt{do\_run}. Compared to the original source, this code is obfuscated. An example of this code is listed in \autoref{fig:lst:prog-delete-listen-ghidra}.
We perform another `Q\&A' session using this code:

\begin{figure}[h!]
\centering
\input{fig/listings/delete-listen-ghidra}
\vspace{-3mm}
\caption{Decompiled (using \texttt{gcc -o1}) \textbf{delete\_listen} from \autoref{fig:lst:prog-delete-listen} (with some reformatting).}
\label{fig:lst:prog-delete-listen-ghidra}
\vspace{-5mm}
\end{figure}

\begin{q_and_a}
\question{What is the purpose of the above code?}
\answer{\xmark}{To create a backdoor on the victim's machine.}

\question{What variable in del\_in\_dir represents the folder to delete files from?}
\answer{\cmark}{param\_1}

\question{What is the format of the message the client sends to the server?}
\answer{\xmark}{The client sends a string of length 0x400 to the server.}

\question{What IP address and port does the server listen on?}
\answer{\cmark}{The server listens on port 0x1f90 on the IP address 192.168.0.1.}
\end{q_and_a}

Despite the increased difficulty of understanding the code\footnote{at least to the human authors}, \davinci{} can still extract useful information in response to the prompts. 
We did not note, however, 
that the exact wording of the question can have a significant impact on the generated answer: when re-wording Q15 into Q17-Q19, the \ac{LLM} \textit{almost} manages to describe the protocol correctly.

\begin{q_and_a}
\question{What is the format of the message the server will process?}
\answer{?}{The server will process a message in the format of a string of characters.}

\question{What commands may the client send to the server?}
\answer{?}{The client may send the following commands to the server: /
    - d: delete a file or directory /
    - q: quit /
    - r: run a file} /

\question{What two commands does this server respond to and what do they do?}
\answer{?}{The server responds to the commands "d" and "q". The "d" command deletes all files in the current directory. The "q" command quits the server.}
\end{q_and_a}

Finally, we can examine the code when it is \textit{stripped} of all identifiers upon compilation. This occurs when compiling with the \texttt{-s} flag with \texttt{gcc}. The code will be the same as listed in \autoref{fig:lst:prog-delete-listen-ghidra} except for the function names.  \texttt{FUN\_0010143a} replaces \texttt{main}, \texttt{FUN\_00101394} replaces \texttt{do\_run}, and \texttt{FUN\_001012c9} replaces \texttt{del\_in\_dir}:

\begin{q_and_a}
\question{What is the purpose of the above code?}
\answer{\xmark}{To create a backdoor on the victim's machine.}

\question{Will any of these functions delete files? If so, which?}
\answer{\cmark}{Yes, the function FUN\_001012c9 will delete files.}

\question{What variable in FUN\_001012c9 represents the folder to delete files from?}
\answer{\cmark}{param\_1}

\question{What is the format of the message the client sends to the server?}
\answer{\xmark}{The client sends a string of length 0x400 to the server.}

\question{What IP address and port does the server listen on?}
\answer{\xmark}{192.168.0.1:54321}
\end{q_and_a}

The quality of the answers suggests the model might be able to deduce useful information from code after compilation and decompilation, even when it is \textit{stripped} of all identifiers. %

\subsubsection{A basic `industrial controller' algorithm}
\label{sec:initial-explore-industry}

\begin{figure}[t!]
\centering
\input{fig/listings/pid-d}
\vspace{-3mm}
\caption{ICS \textbf{pid\_d} program listing (some reformatting).}
\label{fig:lst:prog-pid-d}
\vspace{-5mm}
\end{figure}

Here, we present a similar investigation of using the \ac{LLM} to help us understand a second program: a \ac{PID} controller written in C and listed in \autoref{fig:lst:prog-pid-d}. We call this example \textbf{pid\_d} as the constants are stored in a C \texttt{struct} in dynamic memory.
As we discussed in \autoref{sec:motivation-industrial-re}, extraction or reconstruction of the mathematical formulae and properties used in a given system is invaluable when performing the reverse engineering of a given \acf{ICS}.
We now explore reverse engineering \textbf{pid\_d} while keeping \texttt{temperature} t=$0.00$.

\begin{q_and_a}
\question{What is the purpose of the above code?}
\answer{\cmark}{To calculate the output of a PID controller.}

\question{Given the above code is a PID controller, what is the value of Kp, Ki, and Kd?}
\answer{\cmark}{Kp = 0.7, Ki = 0.1, Kd = 0.2.}

\question{Is there an integrator wind-up guard in the above code? If so, what is its value?}
\answer{\xmark}{Yes, the integral wind-up guard is 1.0.}

\question{Is the output of the PID restricted between any minimum and maximum values? If so, what are they?}
\answer{\cmark}{No, the output is not restricted between any minimum and maximum values.}
\end{q_and_a}

Clearly, the \ac{LLM} continues to show promise in helping with reverse engineering. 
Even though \davinci{} incorrectly claimed that there was a wind-up guard in Q27, it correctly identified the values of $Kp$, $Ki$, and $Kd$.

When we use the same randomization and compilation/decompilation transformations as in \autoref{sec:initial-explore-malware} to create \autoref{fig:lst:prog-pid-d-random} and \autoref{fig:lst:prog-pid-d-ghidra} (in Appendix), consider the responses to the quiz questions. 
Note that (r. src) indicates the randomized source code, \texttt{gcc -o1} indicates compilation by \texttt{gcc} and decompilation by Ghidra, and \texttt{-o1 -s} indicates the same, but with the \texttt{-s} flag to remove all symbols.

\begin{q_and_a}
\question{What is the purpose of the above code?}
\answerM{r. src}{?}{To calculate the value of a function.}
\answerM{\texttt{-o1}}{\cmark}{To calculate the PID values for the PID controller.}
\answerM{\texttt{-o1 -s}}{\xmark}{To print the values of the function $f(x) = x^3 + x^2 + x + 1$}

\question{What mathematical function is encoded in the above code?}
\answerM{r. src}{\xmark}{The above code encodes a linear regression model.}
\answerM{\texttt{-o1}}{\cmark}{The mathematical function is the PID controller.}
\answerM{\texttt{-o1 -s}}{\xmark}{The mathematical function is the function that computes the sine of an angle.}

\question{True or False: The above code implements a PID controller?}
\answerM{r. src}{\cmark}{True.}
\answerM{\texttt{-o1}}{\cmark}{True.}
\answerM{\texttt{-o1 -s}}{\cmark}{True}

\question{True or False: The above code does not implement a PID controller?}
\answerM{r. src}{\cmark}{False. The above code implements a PID controller.}
\answerM{\texttt{-o1}}{\cmark}{False. The above code implements a PID controller.}
\answerM{\texttt{-o1 -s}}{\cmark}{False. The above code implements a PID controller.}

\question{Given the above code is a PID controller, what is the value of Kp, Ki, and Kd?}
\answerM{r. src}{\cmark}{Kp = -0.4830721663176343, Ki = -0.6301845685786949, Kd = -0.96529298213277.}
\answerM{\texttt{-o1}}{\cmark}{Kp = 0x3f333333, Ki = 0x3dcccccd, Kd = 0x3e4ccccd}
\answerM{\texttt{-o1 -s}}{\xmark}{Kp = 0.1, Ki = 0.1, Kd = 0.1}

\question{Is there an integrator wind-up guard in the above code? If so, what is its value?}
\answerM{r. src}{\xmark}{Yes, the integrator wind-up guard is 0.96529298213277.}
\answerM{\texttt{-o1}}{\xmark}{Yes, the integrator wind-up guard is 0x3f333333.}
\answerM{\texttt{-o1 -s}}{\xmark}{Yes, the output of the PID is restricted between -1 and 1.}

\question{Is the output of the PID restricted between any minimum and maximum values? If so, what are they?}
\answerM{r. src}{\xmark}{Yes, the output of the PID is restricted between -1 and 1.}
\answerM{\texttt{-o1}}{\xmark}{The output of the PID is restricted between 0 and 1.}
\answerM{\texttt{-o1 -s}}{\xmark}{Yes, the output of the PID is restricted between -1 and 1.}
\end{q_and_a}

As the code becomes more obscure, the LLM's responses become less helpful, with almost entirely incorrect answers formed when the stripped ($-o1 -s$) version of this code is used. However, worth noting is the success rate of the Q31 and Q32 pair of questions: when directly prompted with the true/false question, \davinci\ was successful in identifying the program (and by testing the negative of the question, we further validated this).

Let us consider Q33 more closely. As the compilation and decompilation process does not affect the order of the parameters in the code (i.e., the common ordering for a PID controller has the parameters in the order Kp $\rightarrow$ Ki $\rightarrow$ Kd), we created a program in which the variables \texttt{Kp}, \texttt{Ki}, and \texttt{Kd} were switched in the function \texttt{pid\_init} (line 12 in \autoref{fig:lst:prog-pid-d}) has a different order of (\texttt{float Ki, float Kd, float Kp}). 

Nonetheless, the LLM could still track the variable contents correctly. That is, the order did not seem to matter and it was still able to answer Q33 correctly.
We also considered the case where rather than asking for the parameter `shortnames' (i.e., Kx---which match the names used in the original source code) we ask for their `proper names', instead posing question Q31 as ``Given the above code is a PID controller, what is the value of the proportional constant, the integral constant, and the derivative constant?.'' It also answered this question correctly.

\section{Exam Time: A Systematic Evaluation}
\label{sec:systematic-evaluation}
While promising, we cannot make a general conclusion on the helpfulness of using the LLM based on solely on the results of \autoref{sec:initial-experiments}. %
Each answer must be judged within the context of its associated program, and deciding if marginal answers should be judged as correct or incorrect is a subjective exercise (e.g., Q19---the two commands are explained, but `the current directory' is not an appropriate designation for the target). Furthermore, quantifying the impact of the different tuning parameters (e.g., \texttt{temperature} and \texttt{top\_p}) is a task that requires well-defined inputs and outputs.

As such, for a comprehensive evaluation, our next experiments quantify the performance of \davinci{} across a set of programs: cybersecurity-related snippets, ICS-relevant algorithms, and real-world malware samples.

\subsection{Experimental Method}

\begin{figure}[t]
    \centering
    \includegraphics[width=0.65\linewidth]{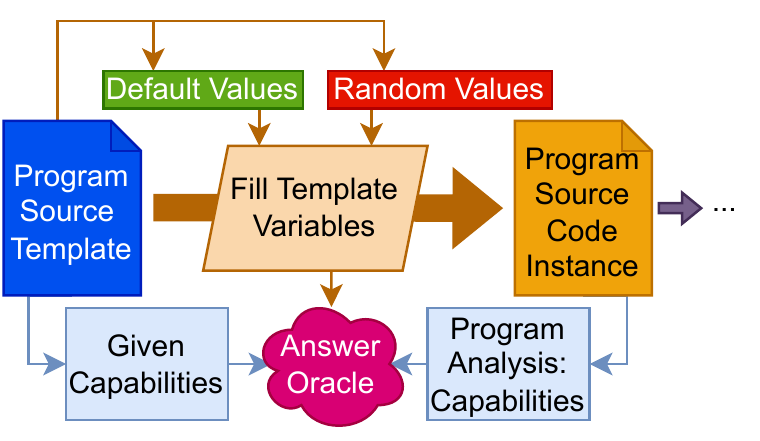}
    \caption{Extending the process illustrated in \autoref{fig:re-proposal} to support \textit{program instances} with known information for extraction (e.g., capabilities, variables).}
    \label{fig:re-templating}
\end{figure}

In this subsection, we detail the set of empirical experiments designed to gauge the performance of \davinci{} when performing `well-defined' reverse engineering tasks. 
The overall aim is to prompt the \ac{LLM} with a series of code snippets and questions about those snippets. 
To prepare the snippets, we create a set of parent ``program source templates'' from which we derive \textit{program source code instances} by randomizing variable/function names, variable values, etc. (\autoref{fig:re-templating}) or by compilation and decompilation (\autoref{fig:re-proposal}). 

For each program source template, we specify its purpose and key information that we want to extract via reverse engineering---this is the set of `correct' answers to the exam. 
When randomizing or compiling-decompiling, we keep track of the correct answers in our \textit{answer oracle}, the marking rubric of this exam. 
We also add to the answer oracle the set of capabilities that each program template implements, as determined by the authors manually and by feeding a compiled instance through the \texttt{capa} tool. 
While creating each source code instance, we generate the corresponding set of questions that can be posed with the source code instance. %

\subsubsection{List of programs}

\label{sec:program-list}
\begin{table}[]
    \centering
    \resizebox{\linewidth}{!}{%
    \newcommand{\STAB}[1]{\begin{tabular}{@{}c@{}}#1\end{tabular}}
\begin{tabular}{@{}llp{5.5cm}p{4cm}@{}}
\toprule
                               & \textbf{Program ID}                                                     & \textbf{Brief Description}                                                                                                                                             & \textbf{Key information to extract}                                                                                \\ \midrule
\multirow{8}{*}{\STAB{\rotatebox[origin=c]{90}{Cybersecurity\hspace{1.44in}}}} & \textbf{delete\_listen}                                                 & Listens on a socket for commands to delete directory contents.                                                                                                & Variables containing files, folder paths; server-client protocol information.                \\
                               & \textbf{delete}                                                         & Simple \textbf{delete\_listen} to delete contents of a statically defined directory.                                                                      & Statically defined folder path.                                                                           \\
                               & \textbf{get}                                                            & Opens and listens on a TCP socket for a file to receive and save.                                                                                             & Variables with file paths; statically defined remote path; destination file name.      \\
                               & \textbf{md5}                                                            & Performs MD5 hashing of a string passed in on the command line.                                                                                               & N/A                                                                                                       \\
                               & \textbf{get\_md5}                                                       & Combination of \textbf{get} and \textbf{md5} programs: the file is downloaded if a given command line argument has a statically defined MD5 hash.                     & Same as \textbf{get}; constant hash value.                                                              \\
                               & \textbf{passwd\_post}                                                   & Reads statically defined file ( /etc/passwd ) and POSTs the contents to a remote URL.                                                                         & Variables containing file paths; statically defined target filename and remote path.                      \\
                               & \textbf{small\_ls}                                                      & A simple \texttt{ls} program to list contents of a specified directory.                                                                                        & N/A                                                                                                       \\
                               & \textbf{xor}                                                            & Performs XOR encryption of a string using a mutating key which starts from a statically defined value.                                            & Initial key; key mutation constant; variables for ciphertext, plaintext, and  key. \\ \midrule
\multirow{6}{*}{\rotatebox[origin=c]{90}{{Industrial Control Systems (ICS)}\hspace{0.14in}}}      & \textbf{pid\_d}                                                         & Proportional-Integral-Derivative (PID) controller whose values are dynamic and kept in a struct in memory.                                       & Kp, Ki, Kd variable names and parameter values; state value variable.                                     \\
                               & \textbf{pid\_s}                                                         & A PID control algorithm where all parameters are static / constant and variables are global.                                                                  & Same as \textbf{pid\_d}.                                                                                         \\
                               & \textbf{fir}                                                            & Finite Impulse Response filter. Parameters are dynamic, kept in a struct.                                                                & Variables for dynamic coefficients, filter state.                                                  \\
                               & \textbf{iir}                                                            & Infinite Impulse Response  filter. Parameters are dynamic, kept in a struct.                                                             & Same as \textbf{fir}.                                                  \\
                               & \textbf{dot\_product}                                                   & Calculate the dot product of two arrays.                                                                                                                      & Variables for dot product.                                                                      \\
                               & \textbf{crc32}                                                          & Calculates the crc32 checksum of a specified file.                                                                                                            & N/A                                                                                                       \\ \midrule
\multirow{5}{*}{\rotatebox[origin=c]{90}{Real Malware\hspace{0.5in}}}  & \textbf{Apachbd}                                                        & A backdoor that provides remote root access in old versions of Apache.                                                                                & N/A                                                                                                       \\
                               & \begin{tabular}[c]{@{}l@{}}\textbf{DblDragon}\\ \textbf{-ranzou}\end{tabular} & Backdoor that bypasses firewalled computers. Daemon runs on target receiving ICMP packets not dropped by firewall. & N/A                                                                                                       \\
                               & \begin{tabular}[c]{@{}l@{}}\textbf{DblDragon}\\ \textbf{-jimmy}\end{tabular}  & Backdoor that bypasses firewalled computers. It communicates with ranzou.                                                        & N/A                                                                                                       \\
                               & \textbf{Dataseg}                                                        & Infector program that modifies a host file's .data segment.                                                                                           & N/A                                                                                                       \\
                               & \textbf{TrojoDaemon}                                                    & Trojan that initiates a process at startup..                                                                                       & N/A                                                                                                       \\ \midrule %
\end{tabular}
    }
    \caption{List of program templates.}
    \label{tab:program-list}
\end{table}

We consider three distinct classes of programs as shown in \autoref{tab:program-list}. For each program, we outline the information that we seek to extract using the \ac{LLM} including purpose of the program as well as key structures/parameters in the program. The first class of program, motivated by \autoref{sec:motivation-malware-re}, consists of small programs like that presented in \autoref{fig:lst:prog-delete-listen}: i.e., cybersecurity-relevant and malware-type codes. The second class, motivated by \autoref{sec:initial-explore-industry}, consist of implementations of algorithms that could be used in \ac{ICS}. Finally, to further examine the \ac{LLM}'s potential usefulness for reverse engineering in `real-world' contexts, we add the third class of programs, which are a set of real malware source code examples from \texttt{vx-underground}~\cite{vxunderground_vxundergroundmalwaresourcecode_2022}.
For simplicity, we focus on the Linux Backdoors, Infectors, and Trojan collections and identified the programs that comprised a single \texttt{C} source file which we were able to compile using \texttt{gcc} without a complex build setup.

\subsubsection{True or false: binary classification questions}
\label{sec:binary-class-method}

\begin{table}[t]
    \centering
    \resizebox{\linewidth}{!}{%
    \begin{tabular}{@{}lll@{}}
\toprule
                               & Short ID              & Capability                                                            \\ \midrule
\multirow{9}{*}{\rotatebox[origin=c]{90}{Cybersecurity}} & Encrypt               & implement an encryption algorithm                                     \\
                               & XOR                   & implement an XOR encryption algorithm                                 \\
                               & AES                   & implement an AES encryption algorithm                                 \\
                               & Delete                & delete file(s)                                                        \\
                               & Network               & interact with a network                                               \\
                               & Downld                & download file(s) from the internet                                    \\
                               & Upload                & upload file(s) to the internet                                        \\
                               & EtcPswd               & read /etc/passwd                                                      \\
                               & MD5                   & implement the MD5 hashing algorithm                                   \\ \midrule
\multirow{9}{*}{\rotatebox[origin=c]{90}{ICS}}           & FIR1                  & implement a FIR filter                                                \\
                               & FIR2                  & implement a Finite Impulse Response filter                            \\
                               & IIR1                  & implement an IIR filter                                               \\
                               & IIR2                  & implement an Infinite Impulse Response filter                         \\
                               & PID1                  & implement a PID controller                                            \\
                               & PID2                  & implement a classic Proportional-Integral-Derivative (PID) controller \\
                               & PI                    & implement a classic Proportional-Integral (PI) controller             \\
                               & PD                    & implement a classic Proportional-Derivative (PD) controller           \\
                               & CSum                  & implement a checksum algorithm                                        \\ \midrule

\multirow{9}{*}{\rotatebox[origin=c]{90}{\texttt{capa-rules} (selected)}}          & DataC-\textgreater{}S & read and send data from client to server                              \\
                               & DataS-\textgreater{}C & receive and write data from server to client                          \\
                               & TCPServer             & start TCP server                                                      \\
                               & ReadFile              & read file on Linux                                                    \\
                               & MoveFile              & move file                                                             \\
                               & WriteFile             & write file on Linux                                                   \\
                               & DNS                   & resolve DNS                                                           \\
                               & ExecuteCmd            & execute command                                                       \\
                               & CreateProc            & create process on Linux                                               \\ \bottomrule
\end{tabular}%
    }
    \caption{List of capabilities used in True/False questions.}
    \label{tab:capabilities-list}
\end{table}

Automatically grading the correctness of the language model outputs can be tricky with open-ended prompts. 
Thus, we prepare a set of true/false questions like those posed in Q31/Q32. 
Taking inspiration from the MITRE ATT\&CK list, we produce a list of capabilities (\autoref{tab:capabilities-list}) and use these to generate a positive question (\textit{True or False: The code above does <capability>.}) and a negative version (\textit{True or False: The code above does not <capability>.}). 
By including both versions of the question, we can gain a sense of whether or not \davinci{} is internally consistent, or has biases towards the true or false answers.

\subsubsection{Short answer: information extraction questions}
\label{sec:info-extraction-method}
Although posing true/false questions provides some insight into an unknown program's capabilities, it does not allow for the practical extraction of \textit{values}.

While one could reasonably pose a true/false question that asked if a given program connected to the internet, determining the remote address using only true/false questions would be extremely time-consuming. 
As such, we also prepare open-ended value questions which are automatically generated during the template instantiation process (\autoref{fig:re-templating}), and are similar to that posed in Q22, Q24, and Q33. Noting that \davinci{} performed better in the preliminary investigation when provided with a context-hint in the question, their general format follows ``Q. Given <some assumption>, what is <some value to extract>?''.

\textbf{Template Tweak.}
During our exploration, we noted that \davinci{} would provide no answer in a small percentage of responses. This was remedied by adding a single `Q\&A example' into the template, after the section break (line 57 in \autoref{fig:lst:prog-delete-listen}) and before the posited question. This example was constant, and read ``Q. What language is the above code written in? <newline> A. C. <newline>''. This avoided introducing bias into any particular answer while ensuring that all \davinci's outputs produced answers.

\subsection{Parameter Tuning}
\label{sec:tuning}

\begin{figure}
    \centering
    \begin{subfigure}[b]{\linewidth}
        \begin{subfigure}[b]{.5\linewidth}
            \centering
            \includegraphics[width=.9\linewidth]{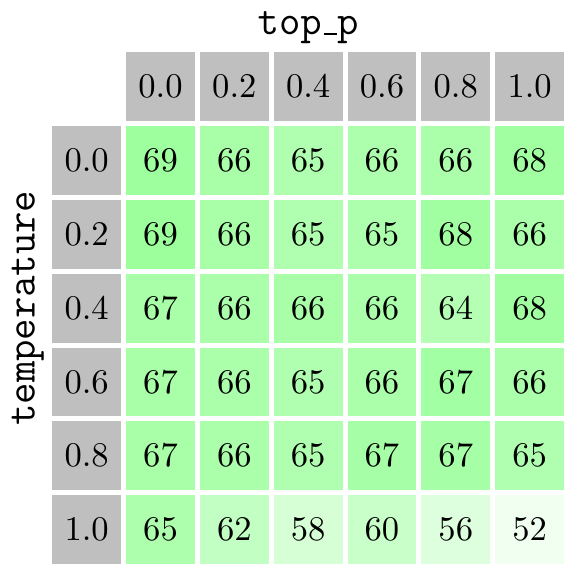}
            \caption{Original source}
        \end{subfigure}%
        \begin{subfigure}[b]{.5\linewidth}
            \centering
            \includegraphics[width=.9\linewidth]{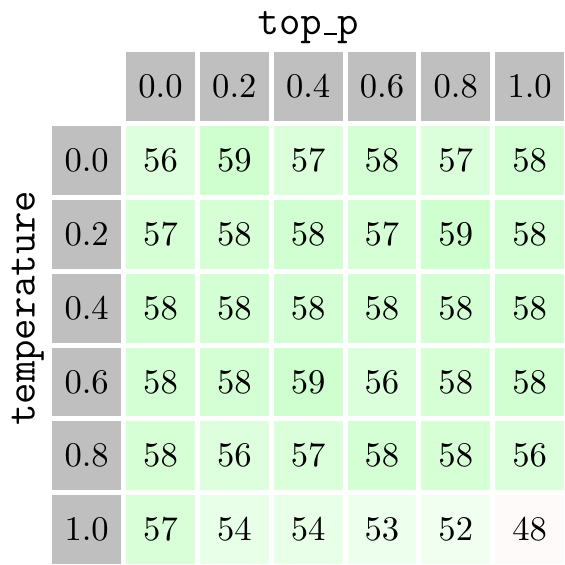}
            \caption{Decompiled \texttt{gcc -o1 -g}}
        \end{subfigure}
    \end{subfigure}
    
    \vspace{1mm}
    \begin{subfigure}[b]{\linewidth}
        \begin{subfigure}[b]{.5\linewidth}
            \centering
            \includegraphics[width=.9\linewidth]{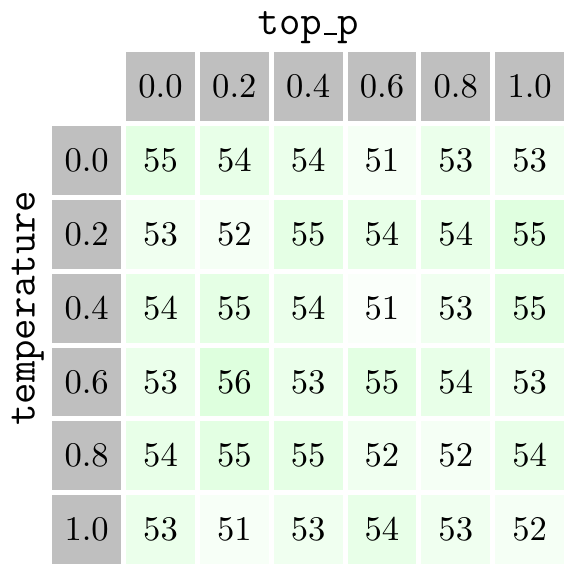}
            \caption{Decompiled \texttt{gcc -o1}}
        \end{subfigure}%
        \begin{subfigure}[b]{.5\linewidth}
            \centering
            \includegraphics[width=.9\linewidth]{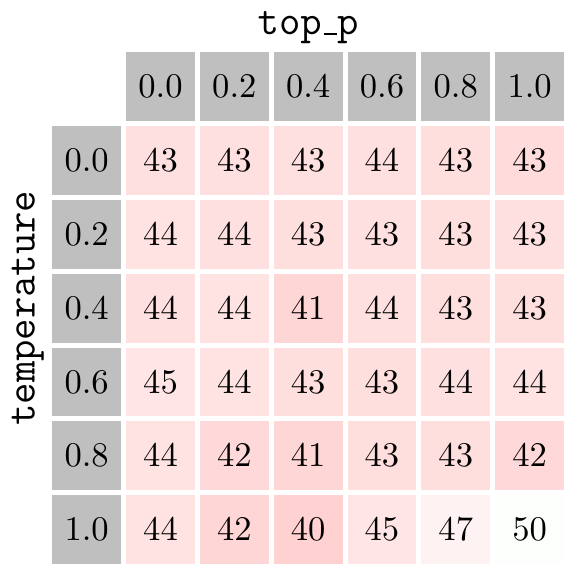}
            \caption{Decompiled \texttt{gcc -o1 -s}}
        \end{subfigure}
    \end{subfigure}

    \vspace{-3mm}
    \caption{Accuracy of True/False classifications when sweeping \texttt{temperature} and \texttt{top\_p}  in steps of $0.2$ for \textbf{pid\_d}.}
    \label{fig:pid-sweep-results}
    \vspace{-3mm}
\end{figure}

\begin{figure*}[t!]
    \centering
    \includegraphics[width=\linewidth]{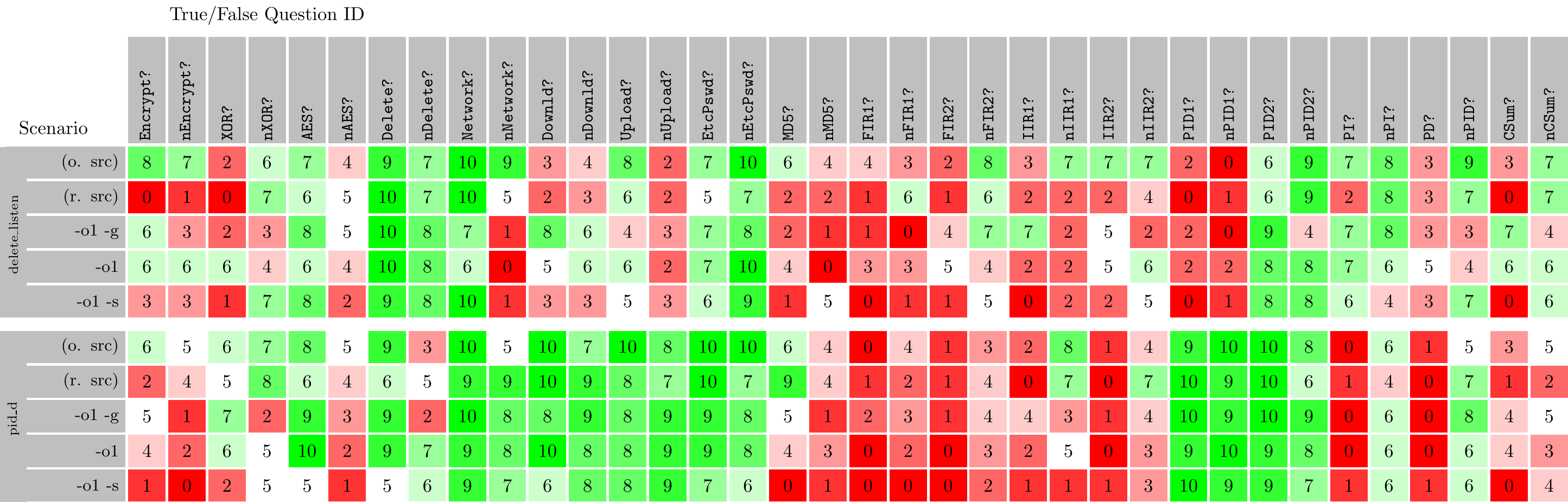}
    \caption{True/False accuracy counts for the \textbf{delete\_listen} and \textbf{pid\_d} programs. All values  are the number of correct answers for that source code configuration and question (maximum: 10).}
    \label{fig:tf-upfront}
\end{figure*}

\begin{figure*}[t!]
    \centering
    \includegraphics[width=\linewidth]{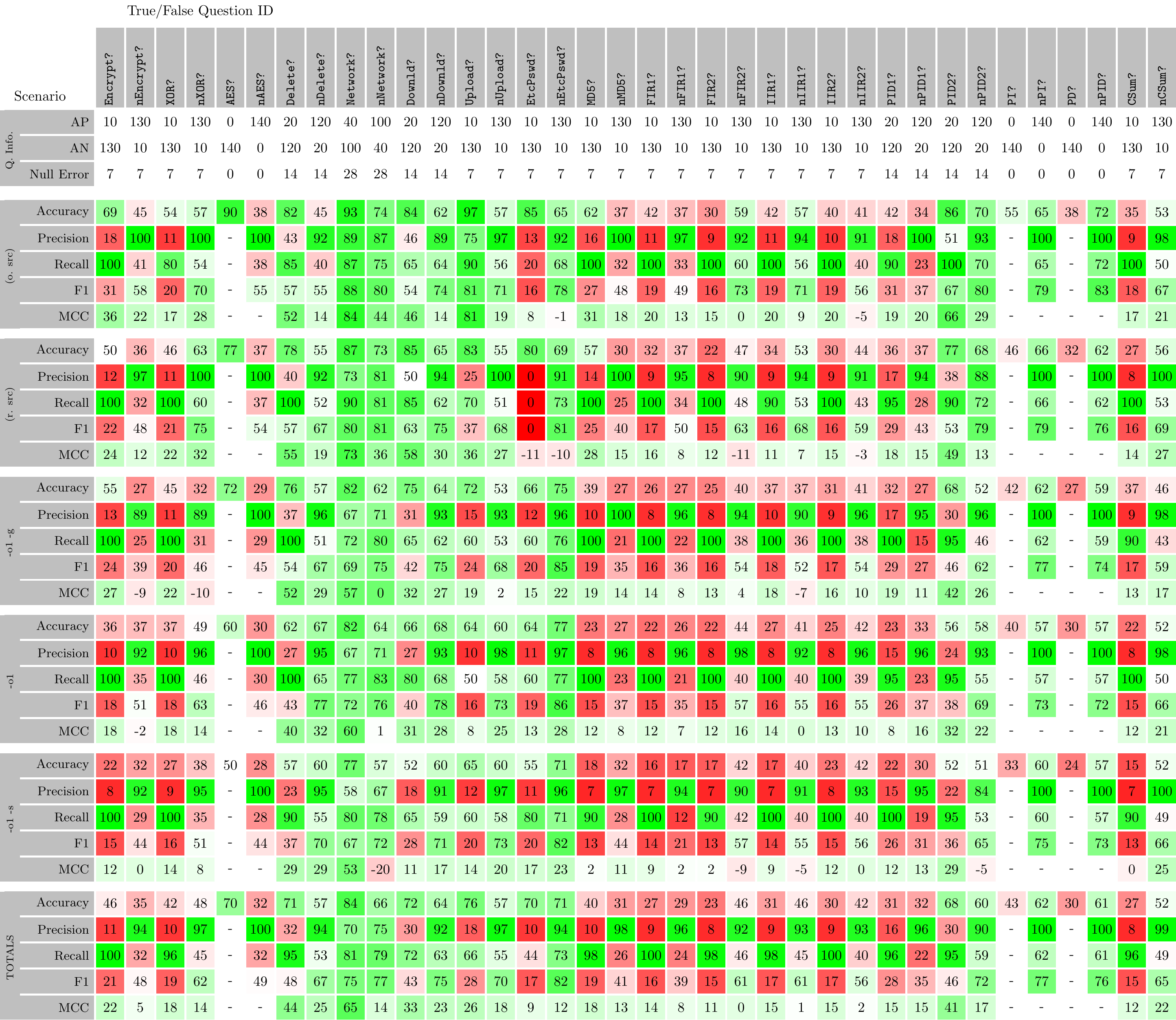}
    \caption{Global True/False statistics across all experiments (\autoref{fig:tf-upfront} (above), and \autoref{fig:tf-general} and \autoref{fig:tf-industrial} (appendix)). \textit{Null Error}, \textit{Accuracy}, \textit{Precision}, \textit{Recall}, and \textit{F1} are presented as rounded percentages (out of 100, bigger values are better). \textit{MCC} is presented rounded and multiplied by 100 (value between -100 and 100, larger/more positive values are better). Cells are marked with "-" when the `actual positive' for that question is 0, as only \textit{accuracy} is meaningful in this case.}
    \label{fig:tf-sums}
\end{figure*}

\begin{figure}
    \centering
    \includegraphics[width=\linewidth]{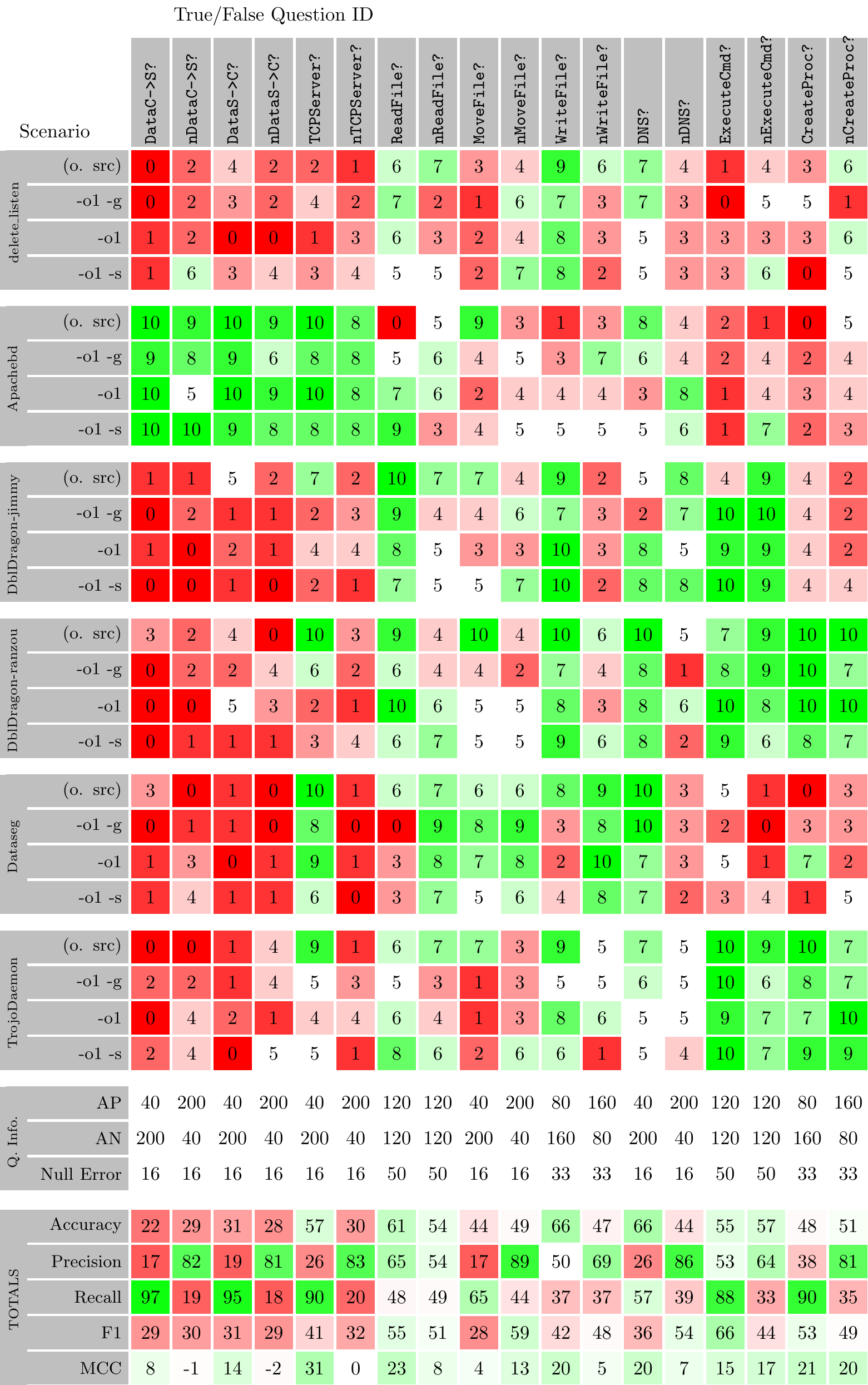}
    \caption{True/False capability Q\&A for real malware.}
    \label{fig:tf-capa}
\end{figure}

As noted in \autoref{sec:initial-experiments}, \davinci\ has two tuning parameters which significantly govern the \ac{LLM}'s outputs: \texttt{temperature}, which affects the probability distribution of each possible token; and \texttt{top\_p}, which controls the sampling for the set of possible tokens at each step of generation. To determine the values that we should select for these parameters, we design an experiment which will conduct a `sweep' across all combinations at step increments of $0.2$ using the experimental methodology laid out in \autoref{sec:binary-class-method}. 
For this purpose, we utilize the \textbf{pid\_d} program, as it is shorter and has fewer features than \textbf{delete\_listen}. 

The results of this experiment are presented in \autoref{fig:pid-sweep-results} for each source code option (the original source code and over Ghidra decompilation of each source code compiled with \texttt{gcc -o1 -g}, \texttt{gcc -o1}, and \texttt{gcc -o1 -s})---to minimise noise during the tuning, no random version was generated or analyzed.
Answers are requested 10 times for each question, i.e., $n=10$. 
In total we asked \underline{51,840 questions}, with \underline{28,630 (55.23\,\%) answered correctly}.
The performance clearly degrades as the code `grows in complexity,' i.e., code is compiled and decompiled with fewer and fewer symbols available.
Across the four experiments, the best parameters are identified as \texttt{top\_p} $=1.0$ and \texttt{temperature} $=0.4$.

\subsection{Grading the True/False Code Purpose Quiz}
\label{sec:grading-tf}

Following the method laid out in \autoref{sec:binary-class-method}, %
we now task \davinci{} with answering the full set of `Q\&A' questions (i.e., positive and negative versions of each capability from \autoref{tab:capabilities-list}) for every program in the corpus (\autoref{tab:program-list}). 
We set \texttt{temperature} $=0.4$ and \texttt{top\_p} $=1.0$ in all trials. 

The results of this experiment are presented thus:
\autoref{fig:tf-upfront} depicts the correct/incorrect counts for the two running examples \textbf{delete\_listen} and \textbf{pid\_d}. For the purposes of identifying the questions in our results, we use the short ID of each capability for the positive question, and the short ID prefixed with an `n' for the negative version. \davinci{} functions better with questions that are `in domain', i.e., the ICS-type true/false questions are answered poorly for the cybersecurity-type program \textbf{delete-listen} and vice versa. Also interesting to note are the cases where the LLM gets only one side of the true/false question correct. For example, \textbf{pid\_d}'s `PI'/`nPI' answers and \textbf{delete\_listen}'s `Upload'/`nUpload' answers. This indicates that under certain scenarios the LLM is answering the same way for both the positive and negative version of a question.
For interest, we include the individual results of the remaining \autoref{tab:program-list} programs in the Appendix \autoref{fig:tf-general} (Cybersecurity) and \autoref{fig:tf-industrial} (\ac{ICS}).

Next, \autoref{fig:tf-sums} presents the combined statistics for all true/false Cybersecurity and ICS questions for each of the experiments. \textit{AP} and \text{AN} reflect the `Actual Positive' (i.e., the number of times this question should have been answered True) and `Actual Negative' (i.e., number of times the question should have been answered False). From this, we compute the \textit{Null Error}, which is the error rate that the \ac{LLM} would get if it always guessed the majority answer. 
As this is a binary classification problem (assume `True'=`Positive', `False'=`Negative'), we present \textit{Accuracy} (the total correct / the total questions), \textit{Precision} (the true positive / the total predicted positive), \textit{Recall} (the true positive / the total actual positive), and the \textit{F1} score, which represents the harmonic mean of \textit{precision} and \textit{recall}. However, as F1 scores can be misleading in cases where the data is imbalanced (as is the case with this corpus), we include a Matthews Correlation Coefficient (MCC)~\cite{chicco_advantages_2020} which will produce a high score (above 0) only when \davinci\ obtains `good' results in each of the four categories. From this we can observe that some questions (e.g., `Network') perform relatively well, while others (e.g., `nFIR2') have results no better than random guessing. In total, we ask \underline{50,400 questions} with \underline{25,048 (49.70\,\%) answered correctly}.

Finally, \autoref{fig:tf-capa} presents a subset of the results for \texttt{capa-rules} capabilities when analyzing real-world malware by comparing the \davinci\ results with the capabilities identified by \texttt{capa}. As with the previous results, some questions (e.g., `TCPServer') are performing better than others (e.g., `nDataS->C'). For \autoref{fig:tf-capa}, we asked \underline{4,320 questions} with \underline{2,036 (47.13\,\%) answered correctly}.

\subsection{Grading the Short Answer Questions} %
\begin{figure}
    \centering
    \includegraphics[width=.9\linewidth]{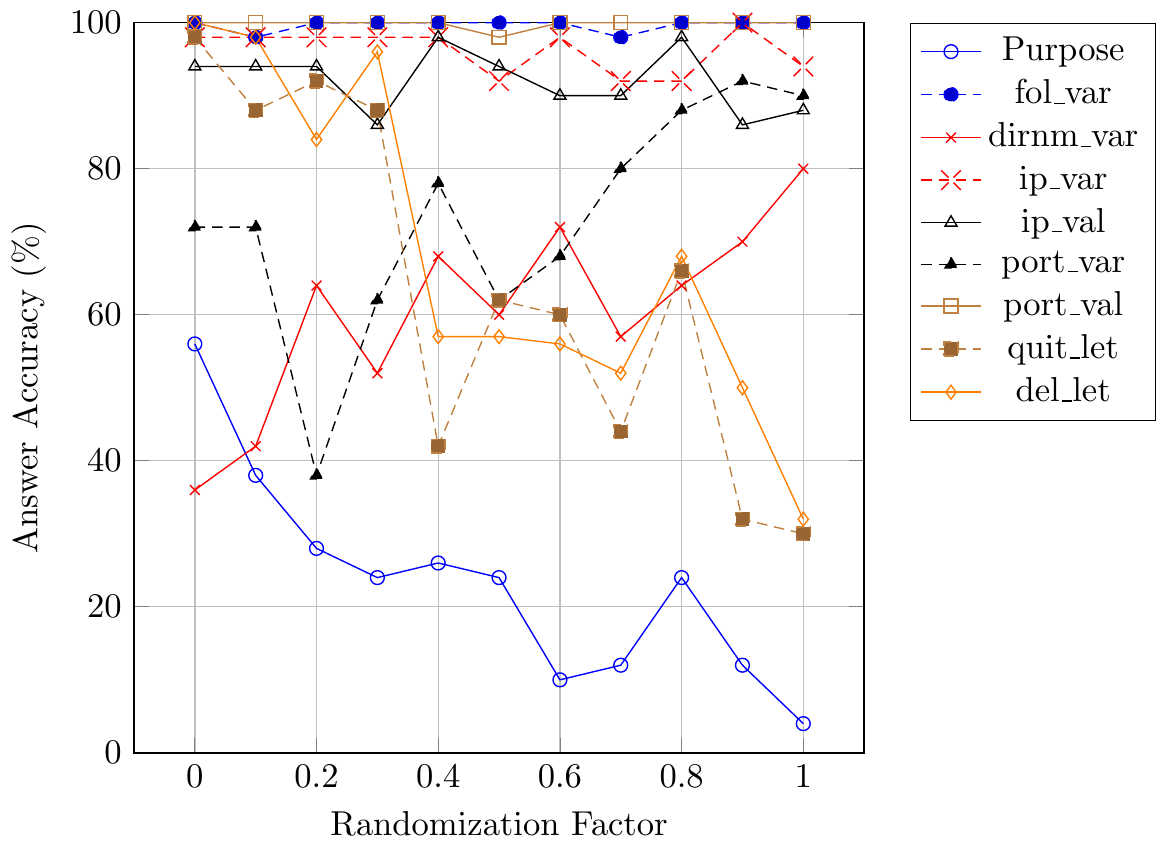}
    \caption{Measuring accuracy in information recovery for open-ended Questions in \textbf{delete\_listen} when increasing level of `code randomization', \texttt{temperature} $=0.4$, \texttt{top\_p} $=1.0$.}
    \label{fig:delete-listen-info-extract-open}
\end{figure}

\begin{figure}
    \centering
    \includegraphics[width=.9\linewidth]{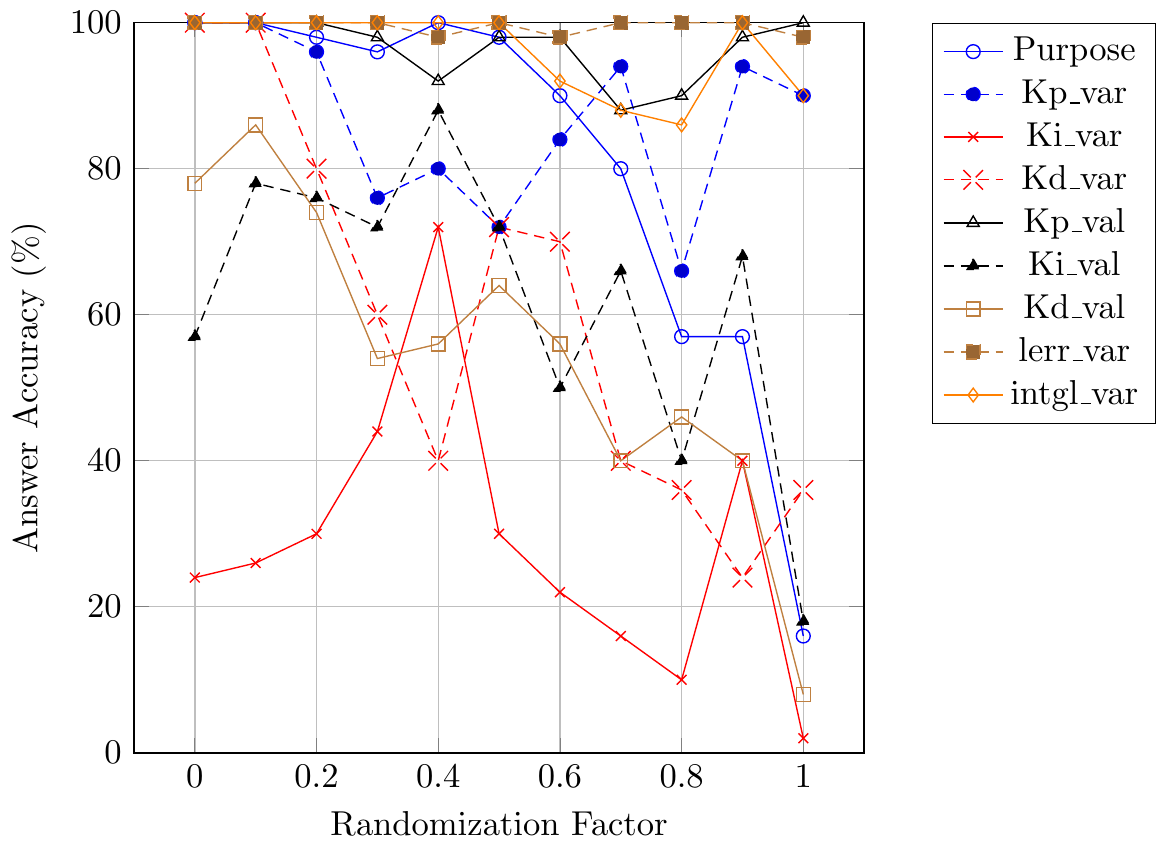}
    \caption{Measuring accuracy in information recovery for Open-ended Questions in \textbf{pid\_d} when increasing the level of `code randomization', \texttt{temperature} $=0.4$, \texttt{top\_p} $=1.0$.}
    \label{fig:pid-d-info-extract-open}
\end{figure}

\label{sec:grading-difficulties}
We now evaluate the LLM using the method as described in \autoref{sec:info-extraction-method}. As the compilation/decompilation pipeline is quite coarse, for this study we only evaluate \davinci's performance over different rates of program randomization.
Using \textbf{delete\_listen} and \textbf{pid\_d}, we create 110 different versions of each, where the first 10 have 0\% randomization, the next 10 have 10\% randomization, the next 20\%, and so on.
Here, the percentage randomization relates to the chance that any given locally defined function name, variable name, or procedural variable content is randomized. We then evaluate the \ac{LLM}'s ability to seek out and identify the key variables and their values.
The results of this investigation are depicted in \autoref{fig:delete-listen-info-extract-open} (\textbf{delete\_listen}) and \autoref{fig:pid-d-info-extract-open} (\textbf{pid\_d}). We also include an additional `Purpose' questions, which pose the following: `Q. In one sentence define the purpose of the above code.' This answer is then examined automatically for a \textit{key word}: in the case of \textbf{delete\_listen}, it is the word ``delete'', in the case of \textbf{pid\_d}, this is the acronym ``PID.'' Overall the results for this experiment are largely as expected given the earlier preliminary experimentation; with the \ac{LLM} performing progressively worse as the randomization increases, especially with the `Purpose' question.

For interest, we also include the capabilities of the \ac{LLM} in answering the True/False question over these difficulty axis as well, depicted in the Appendix in \autoref{fig:delete-listen-info-extract-tf} and \autoref{fig:pid-d-info-extract-tf}. Interestingly, for the True/False questions, the degree of randomization does not seem to have a significantly identifiable trend when considering the average accuracy rate of program identifier questions. 
Across both sets of questions we asked, \underline{29,700 questions} with \underline{17,040 (57.37\,\%) answered correctly}.

\subsection{Key Takeaways}

We set out to understand if \acp{LLM} can aid reverse engineering by identifying key values and variables, code purposes, and capabilities. 
We performed this with \davinci\, which was not trained to summarize code---further, our examples from Cybersecurity and ICS categories were not in its training dataset (as we wrote them for this research). This is a true \textit{zero-shot} setting. As such, it is astonishing that the \ac{LLM} performed as well as it did.
Identifying the capabilities of C programs is challenging even before they are obfuscated (e.g., via decompilation), and yet with no prior exposure, the \ac{LLM} was still answering a narrow majority of questions correctly: (\underline{136,260 questions; 72,754 (53.39\,\%) correct}), including questions that were not just binary True/False but open-ended in nature.
It had some ability to reliably determine program capabilities, especially within the cybersecurity domain programs, and saw some success with real malware samples, where it was able to identify the more straightforward types of capabilities such as file and server interaction.

\section{Course Evaluation: Discussion / Limitations}

While our study provides the first attempt to quantify an \ac{LLM}'s ability to help with reverse engineering, there are a number of limitations that can inform future work. 
We attempted to select a broad range of programs and capabilities; however, there remain many classes of programs that we did not have a chance to explore. 
At present, \davinci{} has a token limit, which limits the size of prompts we can create. 
Given that real-world programs vary in size and complexity, there are situations where the entirety of a program cannot be ingested. 
Prioritizing which parts of a program to include in the prompt (in addition to the snippet of interest) remains an open problem.
In fact, best practice for prompt engineering to coax better responses from an LLM requires further study. 

To glean further insight as to possible reasons why we observed some success when using \davinci{}, which \emph{generates} code, we also explored OpenAI's recently released \emph{embedding} models~\cite{neelakantan_text_2022}. These models transform inputs into $n$-dimensional vectors that can then be clustered and used as features for downstream classification tasks, or compared (e.g., using cosine distance) to compute the similarity between inputs. %
To the best of our understanding, the embedding models are trained using a contrastive training process featuring pairs of (text,code) extracted from docstring comments and their corresponding functions, respectively. 
A pre-trained Transformer model yields an initial embedding for each component using  models such as GPT-3 (for text, denoted here as $E_T$) and Codex (for code, denoted $E_C$) by extracting the hidden state of the last layer for the final token in the input. These initial embeddings are not aligned (i.e., the vector for a code snippet will not be close to the vector for its documentation), %
so a contrastive objective optimizes the weights of the embedding models $E_T$ and $E_C$ to minimize the distance between positive samples (correctly paired text and code snippets %
while maximizing the distance between negative samples. %

The result of this process is a pair of aligned embedding models that represent natural language and code in the same space. We used the larger of OpenAI's two publicly available code embedding models, \babbage, to gain some intuition about Codex's ability to understand code semantics from \emph{decompiled} code. Given that the code embedding model is derived from Codex, we assume that relationships identified with the embeddings API will likely also hold in the Codex space.

\begin{figure}[t]
\centering
\begin{subfigure}{0.48\columnwidth}
\centering
\includegraphics[width=\textwidth]{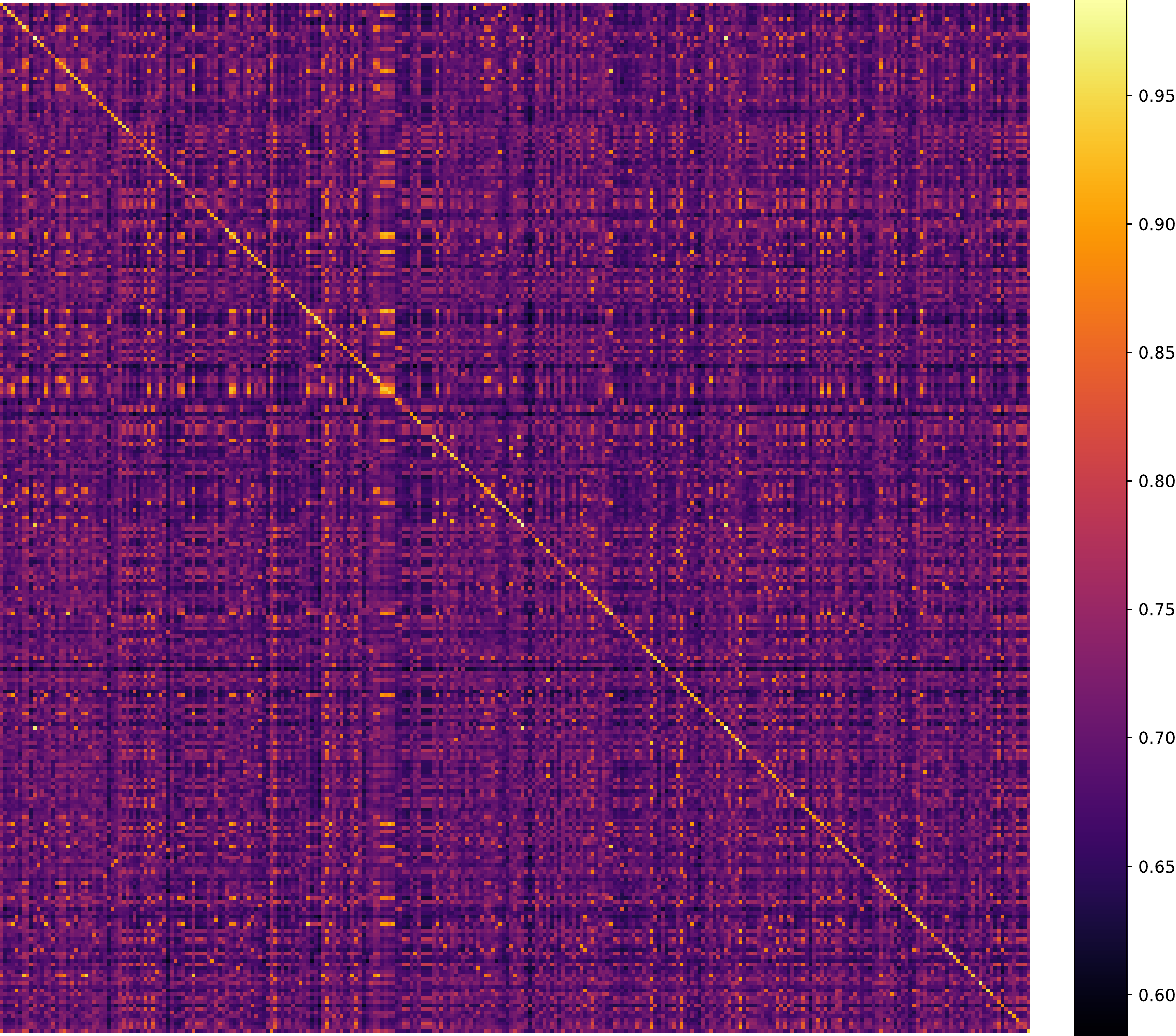}
\caption{Confusion matrix for \texttt{ls} with debug information.}
\label{fig:lsembed_dbg}
\end{subfigure}
\begin{subfigure}{0.48\columnwidth}
\centering
\includegraphics[width=\textwidth]{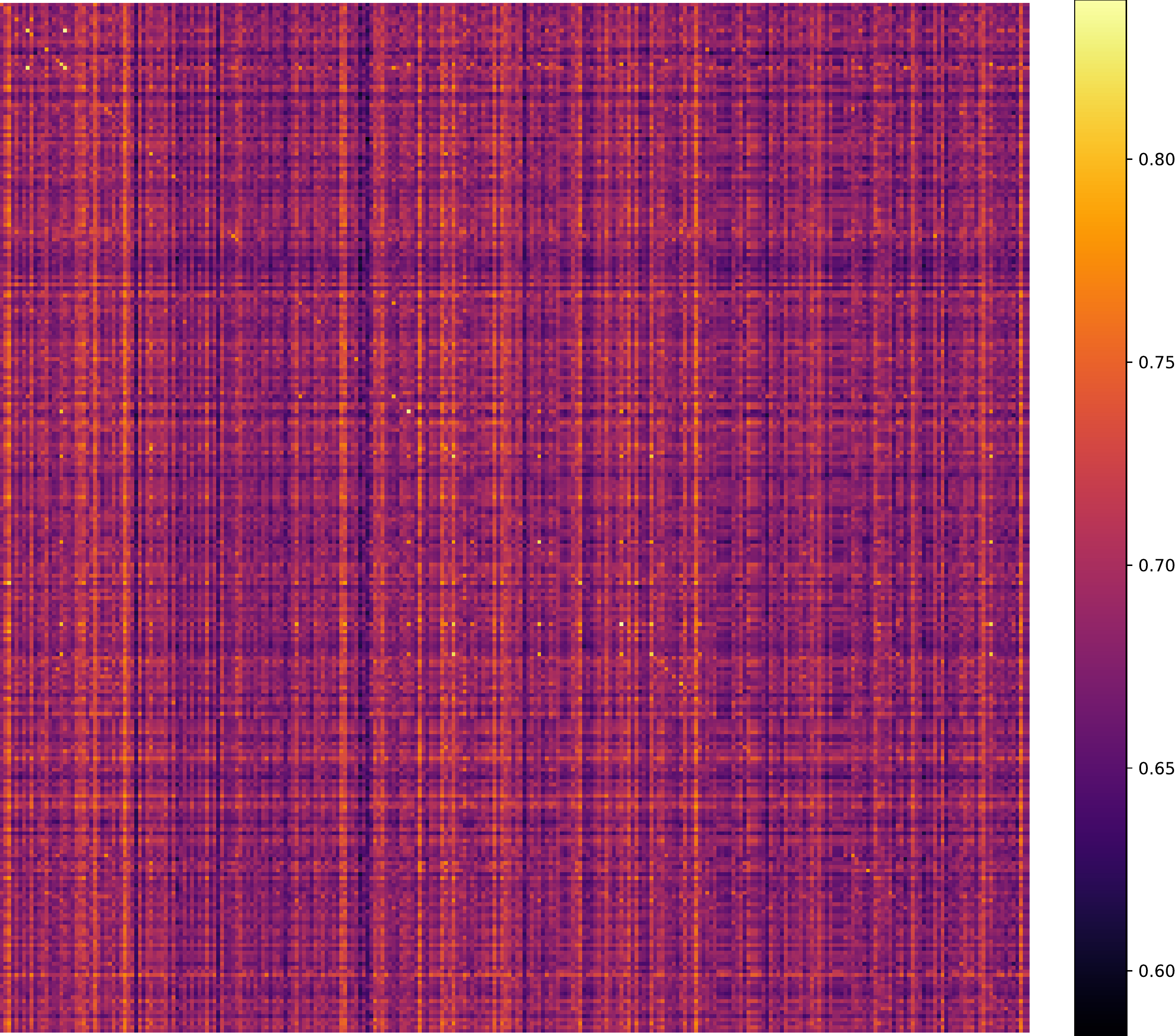}
\caption{Confusion matrix for \texttt{ls} with debug symbols stripped.}
\label{fig:lsembed_strip}
\end{subfigure}
\caption{Confusion matrix heatmap of embedding similarities for original vs. decompiled source across 279 functions from Coreutils \texttt{ls} with debug symbol information (\autoref{fig:lsembed_dbg}) and on a stripped binary (\autoref{fig:lsembed_strip}).}
\label{fig:lsembed}
\end{figure}

\autoref{fig:lsembed} shows a heatmap of cosine distances between the \babbage\ embeddings of 279 functions in the Coreutils version of \texttt{ls} for both the original source code of the function (along the x-axis) and the Ghidra decompilation of its compiled counterpart (y-axis). We can see that when debug symbols are present (\autoref{fig:lsembed_dbg}), the Codex-based embedding of a source function is generally much closer to the embedding of its decompilation than to the decompiled version of any other function in the binary (as indicated by the bright diagonal line running from top-left to bottom-right). After stripping away debug symbols (\autoref{fig:lsembed_strip}), this signal is much weaker, but Codex is still somewhat capable of identifying a function from its decompilation (the diagonal is still visible, albeit harder to see amongst the noise). We take this as a promising indication that Codex is not merely relying on function and variable names, and might be able to extract a function's structure or even semantics to some degree.

\section{Reading List: Related Work\label{sec:relatedwork}}

In this work, our investigation focused on eliciting information from code that could be of interest to a reverse engineer; there are several areas of prior work that attempt similar feats in a broader software engineering context. 

For instance, there have been numerous investigations into techniques for code summarization, from early approaches using text retrieval~\cite{haiduc_use_2010} through to more recent \ac{ML}-based approaches that are often framed as a Neural Machine Translation (NMT) problem (e.g., see recent works by Mastropaolo et al.~\cite{mastropaolo_studying_2021} and Haque et al. \cite{haque_improved_2020}). 
Generally, deep learning approaches for code summarization try to infer information that is not explicitly in the code, being trained on code/comment pairs (such as LeClair and McMillan's dataset~\cite{leclair_recommendations_2019}) and evaluated on regular (i.e., not decompiled) source code. In contrast to these works, we investigated the ability of an \ac{LLM} to explain code that is not as descriptive as the original human-authored code (as a result of stripping/decompilation). 

In reverse engineering, several ML-driven approaches aim to support binary analysis in task-specific ways. %
For example, neural networks have been used for predicting function names~\cite{david_neural_2020, patrick-evans_probabilistic_2020}, function type signatures~\cite{chua_neural_2017}, improving disassembly~\cite{pei_xda_2021}, and end-to-end decompilation~\cite{fu_coda_2019}. 
In contrast, Codex and other LLMs are not specialized. 
We do not claim that LLMs can supplant prior work; we were interested only in characterizing how helpful an LLM, trained on vast quantities of code, can be.  

In the ICS domain, prior works have addressed decompilation \cite{lv_technique_2017} of binaries from embedded devices such as Programmable Logic Controllers (PLCs) and extraction of parameters and logic information. Symbolic execution and program slicing techniques have been applied for automated detection of vulnerabilities such as authentication bypass \cite{shoshitaishvili2015firmalice,shoshitaishvili2016state}. In \cite{keliris_icsref_2019}, a framework was developed to reconstruct the control flow graph, match against known functions, and recover function call arguments from PLC binaries compiled using the widely used CODESYS and applied to extract information such as PID gains. While our Codex-based approach is more analogous to source/binary static analysis approaches, it is to be noted that in scenarios where an appropriate execution environment is available, some of the information extraction goals (e.g., extracting parameters such as PID gains) can also be addressed through methods such as instrumented fuzzing and ICS network traffic analysis \cite{tychalas_icsfuzz_2021,qasim_control_2020}.

\section{Conclusions}

\acp{LLM} have demonstrated numerous zero-shot capabilities across domains such as language translation, program generation, and code explanation. We set out to discover if these capabilities extended to the reverse engineering domain, and found that there is a definite potential. When interacting with a leading \ac{LLM} in a `Q\&A' pop-quiz like format, both binary-classification and open-ended questions could be answered correctly in at least a narrow majority of cases. However, while promising, further maturation of the technologies is required before they could be reliably applied for reverse engineering in a wide range of contexts. Such approaches could involve integration of additional training data (e.g., decompiled codes), or the usage of obfuscation tools. In the future, we think \acp{LLM} could be useful in this domain, but for now, they might need to go back to class.

\vspace{3mm}
\noindent \textbf{Acknowledgments}  %
to OpenAI for providing API access to their language models.

\vspace{3mm}
\noindent \textbf{Availability:}
All source code, scripts, and OpenAI \davinci\ responses are recorded and made available at: \url{https://doi.org/10.5281/zenodo.5949075}.

\clearpage
\bibliographystyle{plainurl}
\bibliography{bib/benhamram,bib/refs}

\clearpage
\appendix

\section*{Appendix: Additional Figures}

\begin{figure}[h]
\centering
\input{fig/listings/pid-d-random}
\vspace{-3mm}
\caption{Randomized \textbf{pid\_d} program (original in \autoref{fig:lst:prog-pid-d}).}
\label{fig:lst:prog-pid-d-random}
\end{figure}

\begin{figure}[h]
\centering
\input{fig/listings/pid-d-ghidra}
\vspace{-3mm}
\caption{Decompiled (from \texttt{gcc -o1}) \textbf{pid\_d} program (original in \autoref{fig:lst:prog-pid-d}). Slightly reformatted here for clarity.}
\label{fig:lst:prog-pid-d-ghidra}
\end{figure}

\begin{minipage}{\columnwidth}
    \centering
    \includegraphics[width=.8\linewidth]{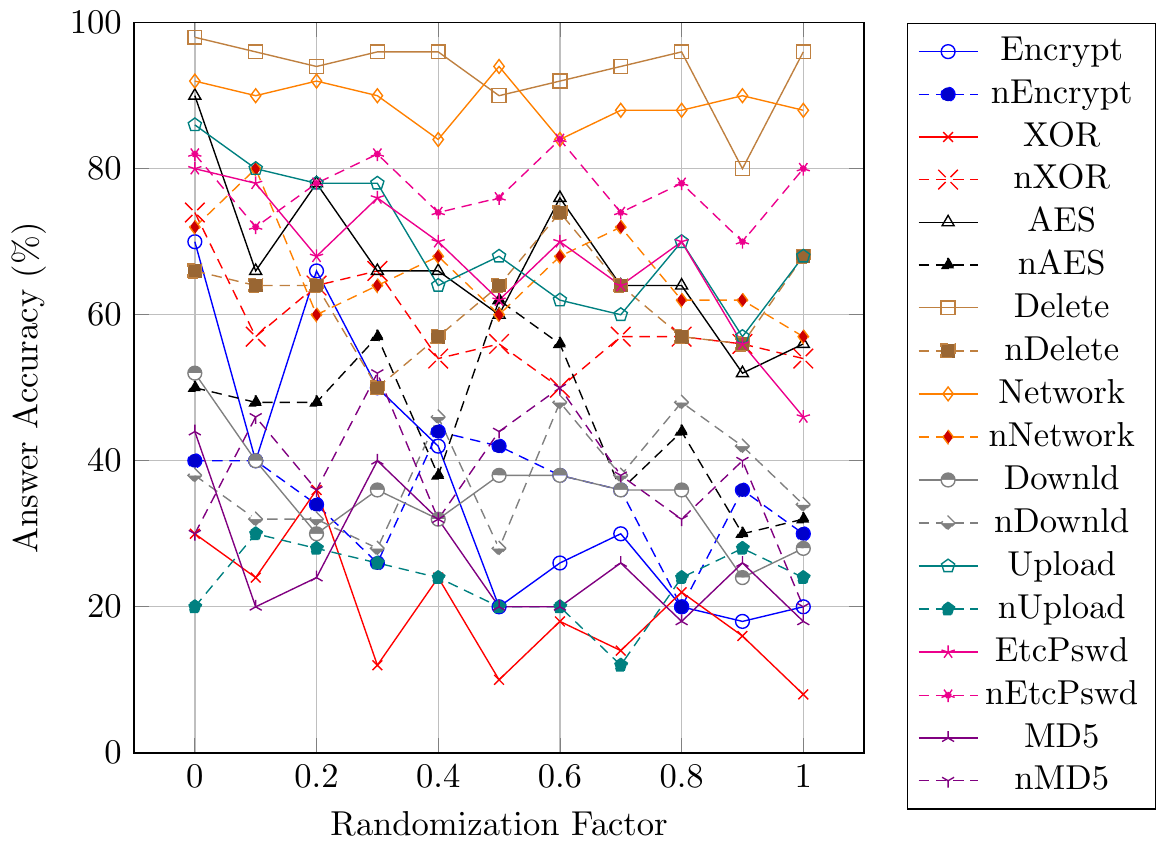}
    \vspace{-5mm}
    \captionof{figure}{Measuring accuracy in information recovery for True/False Questions in \textbf{delete\_listen} when increasing the level of `code randomization', \texttt{temp.} $=0.4$, \texttt{top\_p} $=1.0$.}
    \label{fig:delete-listen-info-extract-tf}
    \vspace{-5mm}
\end{minipage}

\begin{figure}[]
    \centering
    \includegraphics[width=.8\linewidth]{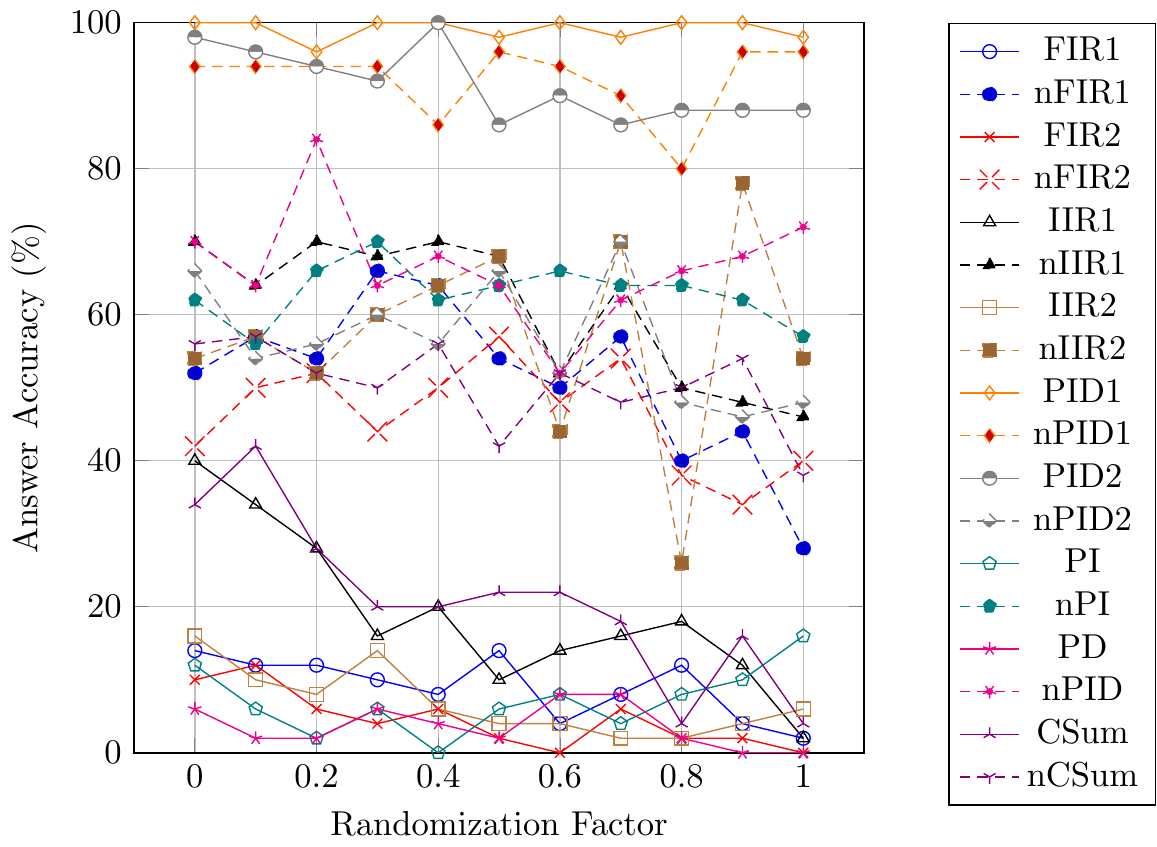}
    \vspace{-5mm}
    \caption{Measuring accuracy in information recovery for True/False Questions in \textbf{pid\_d} program when increasing the level of `code randomization', \texttt{temp.} $=0.4$, \texttt{top\_p} $=1.0$.}
    \label{fig:pid-d-info-extract-tf}
    \vspace{-5mm}
\end{figure}

\begin{figure*}[h]
    \centering
    \includegraphics[width=\linewidth]{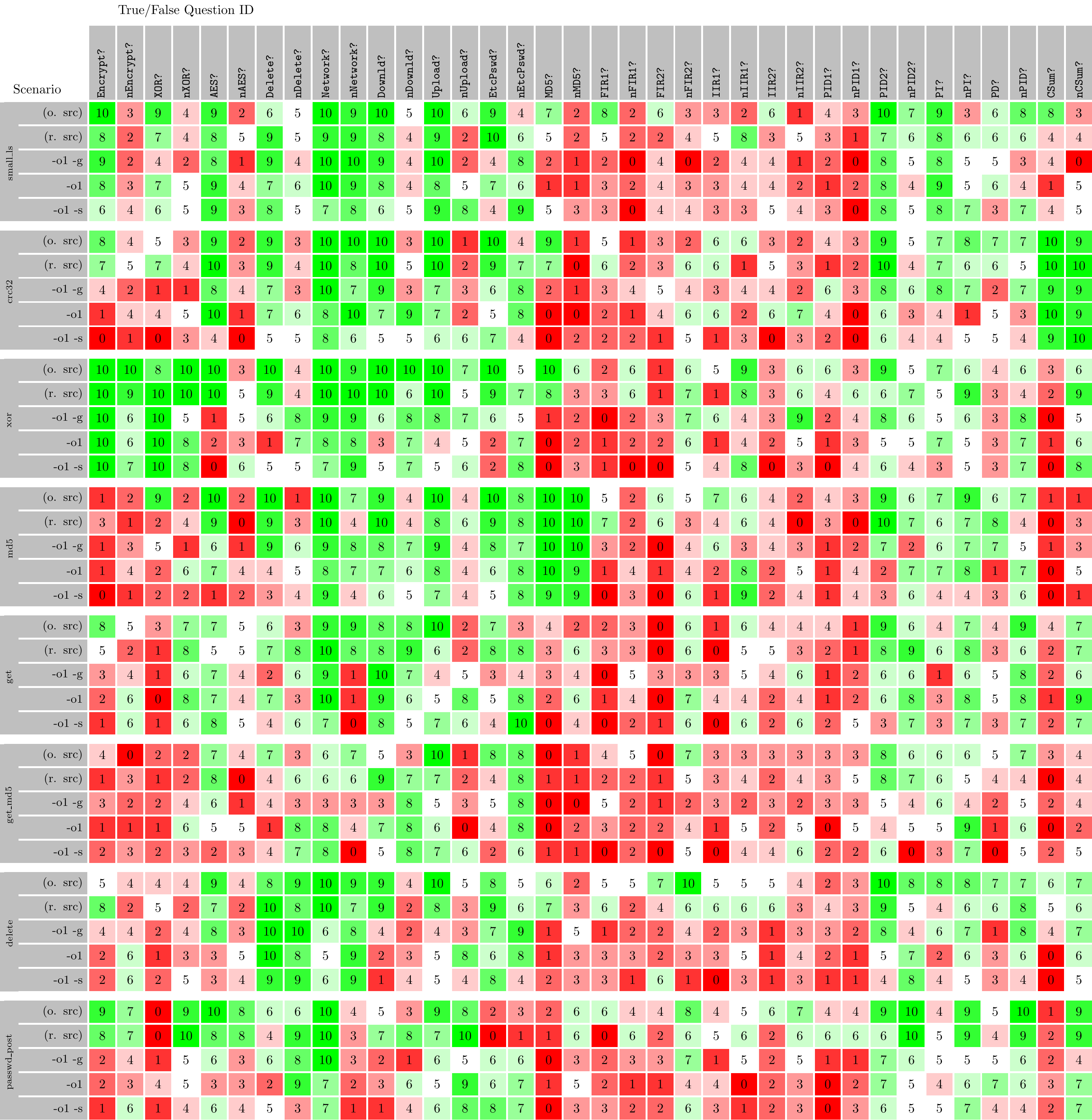}
    \caption{True/False heatmaps for `Cybersecurity' programs. Values are number of correct answers for that question (max: 10).}
    \label{fig:tf-general}
\end{figure*}

\begin{figure*}[h]
    \centering
    \includegraphics[width=\linewidth]{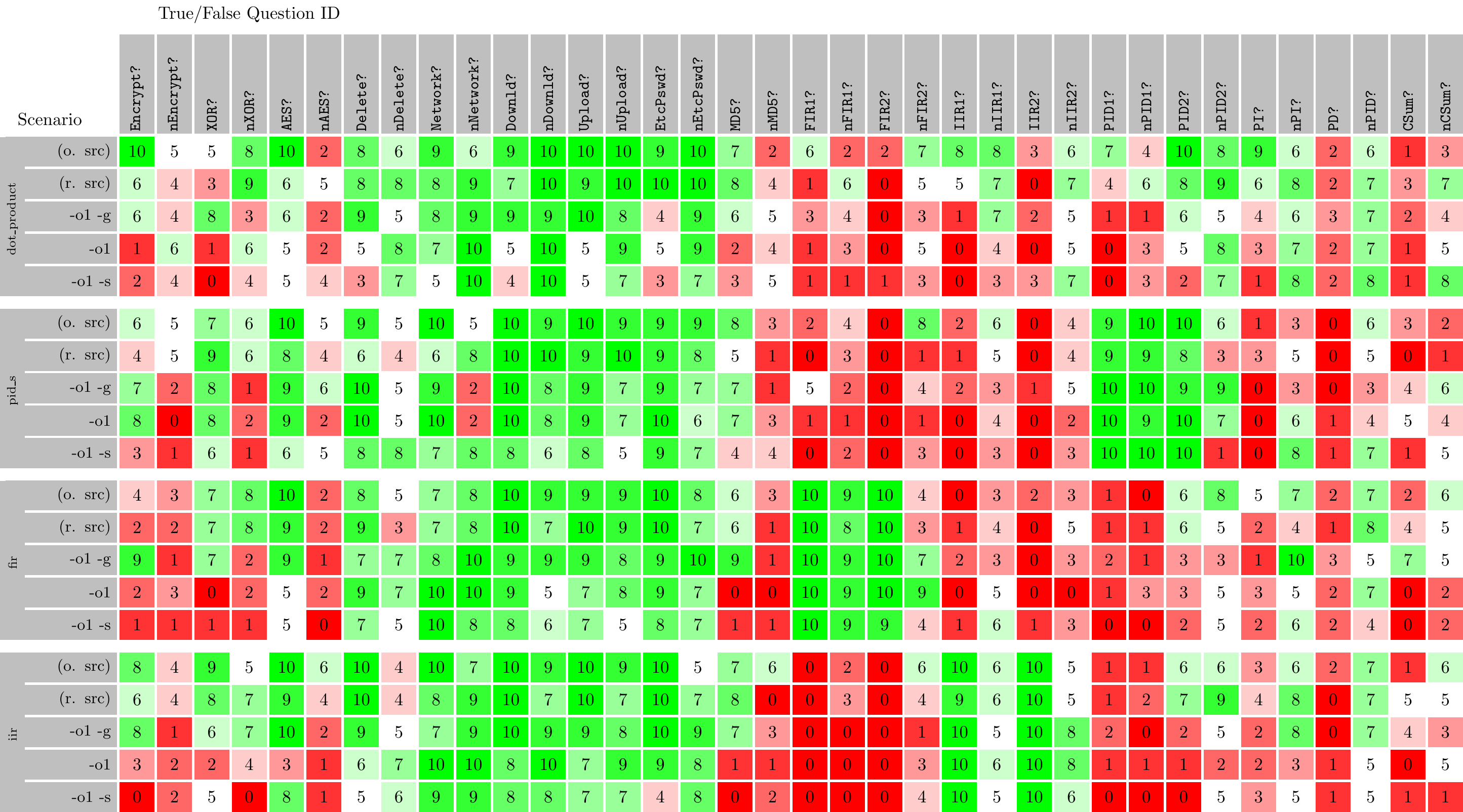}
    \caption{True/False heatmaps for ICS-type programs. Values are number of correct answers for that question (max: 10).}
    \label{fig:tf-industrial}
\end{figure*}

\end{document}